\documentclass[sigconf,screen]{acmart}

\AtBeginDocument{%
  \providecommand\BibTeX{{%
    \normalfont B\kern-0.5em{\scshape i\kern-0.25em b}\kern-0.8em\TeX}}}

\setcopyright{none}
\copyrightyear{2023}
\acmYear{2023}

\acmConference[VAT-HRI '23]{Workshop on Variable Autonomy for Human-Robot Teaming at the ACM/IEEE International Conference on Human-Robot Interaction}{March 13, 2023}{Stockholm, SE}
\acmBooktitle{VAT-HRI '23: Workshop on Variable Autonomy for Human-Robot Teaming at the ACM/IEEE International Conference on Human-Robot Interaction, March 13, 2023, Stockholm, SE} 

\usepackage[autostyle=true,english=american]{csquotes}
\usepackage{acro}
\DeclareAcronym{UI}{short = UI, long = user interface}
\DeclareAcronym{GUI}{short = GUI, long = graphical user interface}
\DeclareAcronym{TLX}{short = NASA TLX, long = NASA-Task Load Index}
\DeclareAcronym{RTLX}{short = NASA Raw-TLX, long =NASA Raw-Task Load Index}
\DeclareAcronym{ER}{short = ER, long = error rate}
\DeclareAcronym{TCT}{short = TCT, long = task completion time}
\DeclareAcronym{HCI}{short = HCI, long = Human-Computer Interaction}
\DeclareAcronym{UX}{short = UX, long = user experience}
\DeclareAcronym{RMSE}{short = RMSE, long = root mean squared error}
\DeclareAcronym{HMD}{short = HMD, long = Head-Mounted Display}
\DeclareAcronym{CNN}{short = CNN, long = Convolutional Neural Network}
\DeclareAcronym{FOV}{short = FoV, long = field of view}
\DeclareAcronym{HRC}{short = HRC, long = Human-Robot Collaboration}
\DeclareAcronym{HRI}{short = HRI, long = Human-Robot Interaction}
\DeclareAcronym{ANOVA}{short = ANOVA, long = analysis of variance}
\DeclareAcronym{RMANOVA}{short = RM-ANOVA, long = Repeated Measures Analysis of Variance}
\DeclareAcronym{JND}{short = JND, long =just-noticeable difference}
\DeclareAcronym{SUS}{short = SUS, long =system usability scale}
\DeclareAcronym{CSCW}{short = CSCW, long = computer-supported cooperative work}
\DeclareAcronym{CAD}{short = CAD, long = computer-aided design}
\DeclareAcronym{MR}{short = MR, long = Mixed Reality}
\DeclareAcronym{AR}{short = AR, long = Augmented Reality}
\DeclareAcronym{AV}{short = AV, long = Augmented Virtuality}
\DeclareAcronym{VR}{short = VR, long = Virtual Reality}
\DeclareAcronym{SAR}{short = SAR, long = Spatial Augmented Reality}
\DeclareAcronym{ADLs}{short = ADLs, long = Activities of Daily Living}
\DeclareAcronym{LED}{short = LED, long = Light-Emitting Diode}
\DeclareAcronym{DoF}{short = DoF, long = Degree-of-Freedom}
\DeclareAcronym{DoFs}{short = DoFs, long = Degrees-of-Freedom}
\DeclareAcronym{HHC}{short = HHC, long = Human-Human Collaboration}
\DeclareAcronym{AI}{short = AI, long = Artifical Intelligence}
\DeclareAcronym{QUEAD}{short = QUEAD, long = Questionnaire for the Evaluation of Physical Assistive Devices}
\DeclareAcronym{TiA}{short = TiA, long = Trust in Automation Questionnaire}
\DeclareAcronym{TOR}{short = TOR, long = Take-Over-Request}
\DeclareAcronym{ADMC}{short = ADMC, long = Adaptive DoF Mapping Controls}

\begin{document}

\title{Understanding Shared Control for Assistive Robotic Arms}


\author{Kirill Kronhardt}
\orcid{0000-0002-0460-3787}
\email{kirill.kronhardt@.w-hs.de}
\affiliation{
    \institution{Westphalian University of Applied Sciences}
    \city{Gelsenkirchen}
    \country{Germany}
}

\author{Max Pascher}
\orcid{0000-0002-6847-0696}
\email{max.pascher@w-hs.de}
\affiliation{
    \institution{Westphalian University of Applied Sciences}
    \city{Gelsenkirchen}
    \country{Germany}
}
\affiliation{
    \institution{University of Duisburg-Essen}
    \city{Essen}
    \country{Germany}
}

\author{Jens Gerken}
\orcid{0000-0002-0634-3931}
\email{jens.gerken@w-hs.de}
\affiliation{
    \institution{Westphalian University of Applied Sciences}
    \city{Gelsenkirchen}
    \country{Germany}
}


\begin{abstract}
Living a self-determined life independent of human caregivers or fully autonomous robots is a crucial factor for human dignity and the preservation of self-worth for people with motor impairments.
Assistive robotic solutions -- particularly robotic arms -- are frequently deployed in domestic care, empowering people with motor impairments in performing \ac{ADLs} independently. However, while assistive robotic arms can help them perform ADLs, currently available controls are highly complex and time-consuming due to the need to control multiple \ac{DoFs} at once and necessary mode-switches. This work provides an overview of shared control approaches for assistive robotic arms, which aim to improve their ease of use for people with motor impairments. We identify three main takeaways for future research: \textit{Less is More}, \textit{Pick-and-Place Matters}, and \textit{Communicating Intent}.
\end{abstract}





\maketitle

\section{Introduction}
People with motor impairments often require caregivers to help them perform \ac{ADLs} such as eating or drinking \citep{Martinsen.2008}. This near-constant need for care causes a desire for more autonomy \citep{Pascher.2021recommendations}. There are multiple ways to accomplish a higher level of autonomy for people with motor impairments using assistive robots. One approach often seen in research is the use of autonomous assistive robots \citep{Kyrarini.2021survey}. In these autonomous systems, robots perform \ac{ADLs} automatically while their users only provide high-level input or interaction, e.g. issuing a command to stop the robot from moving to prevent a collision. However, such autonomous systems can cause more stress in their users than manual control methods. In a study comparing manual control and autonomous behavior of a robotic arm, \citet{Pollak.2020} found significantly reduced stress symptoms for the manual mode. A similar comparative study by \citet{Kim.2012} revealed that while autonomous robots reduce the effort required to perform \ac{ADLs}, there were no significant differences in satisfaction levels between autonomous and manual controls. Furthermore, \citet{Kim.2012} suggest that people with motor impairments experience autonomous robots as another external agent to rely on, which does not result in a higher experienced level of autonomy. These findings align with more general research on autonomous robots by \citet{zlotowski2017can}, where participants were found to generally have more negative feelings towards autonomous robots than towards non-autonomous ones.

While research shows that completely autonomous robots are unfit to mitigate the experienced lack of autonomy people with motor impairments face, completely manual controls also encompass considerable drawbacks. These drawbacks are especially apparent with assistive robotic arms, which can often move with six or more \ac{DoFs}. In contrast, input devices for manual controls accessible to people with motor impairments, such as two-dimensional joysticks, offer much fewer \ac{DoFs}~\citep{Maheu.2011,Prattico.2021}. This discrepancy necessitates a mapping of movement \ac{DoFs} -- or output-\ac{DoFs} -- of the robot arm to input-\ac{DoFs} of the input device, as well as mode switches to change the active mapping. These mode switches lead to an unsatisfying experience, requiring too many inputs and too much time to execute \ac{ADLs} to be useful for people with motor impairments \citep{eftring1999technical, tijsma2005framework, Herlant.2016modeswitch}. They also require high concentration as remembering the robot's current mode is difficult, and little-to-no visualization of the active mode is provided \citep{Herlant.2016modeswitch}. In particular, users experience these mode switch-based controls as cumbersome and time-consuming. In interviews conducted by \citet{Herlant.2016modeswitch}, three daily users of the \emph{Kinova Jaco} assistive robot arm reported abandoning elementary tasks like eating due to the frequency of mode switches, instead relying on caregivers. \citet{Herlant.2016modeswitch} also found that about 18\% of task execution time when controlling a robotic arm using mode switching is spent switching modes instead of performing movements. These findings are in line with user trials of the \emph{Manus} assistive robotic arm conducted by \citet{eftring1999technical}, where the controls were largely experienced as \enquote{too slow} and requiring \enquote{too many commands}. Therefore, research should focus on improving the usability and ease of control for assistive robotic arms. 

To tackle the dilemma between the difficulty of manual controls and the lack of autonomy experienced when collaborating with a fully autonomous robot, different approaches along the continuum of shared control methods were proposed \citep{Goldau.2021petra, Kronhardt.2022adaptOrPerish,Herlant.2016modeswitch,tsui2011want,ezeh2017,quere2020,jain2015}.
The concept of shared control has great potential to design communication and control between human and robot~\cite{Abbink2018}. For assistive robotic arms, however, it should also lead to meaningful interaction for the user and a feeling of being in control. Shared control systems for assistive robotic arms can give users a feeling of independence while improving ease of use compared to manual controls. This paper provides an overview of recent advances and different approaches to realize such shared control mechanisms for assistive robotic arms.

\section{Shared Control for Assistive Robotic Arms}
\label{sec:current-shared-controls}

Shared control systems exist on a spectrum of autonomy \citep{ERDOGAN2017282}. This spectrum consists of systems that lean heavily towards manual control, only adjusting the results of the users' input slightly, as well as systems where users mainly give high-level input commands for the robot to perform. 
An example of such a system was developed by \citet{tsui2011want}, where users only indicate an object to be picked, and a robotic arm executes the picking motion autonomously. 

As we highlighted, for assistive robotic arms, shared control systems should allow users to feel more independent than systems that rely more heavily on the robot's autonomy allow them to feel. Therefore, the shared control systems used for assistive robotic arms should lean more toward user autonomy. A fundamental and generalizable use case for any robotic arm that is flexible enough to support many different tasks is the pick-and-place operation. In an analysis of 11 hours of lifelogging video data of non-impaired people, \citet{Petrich.2022ADL} found that picking and placing objects are highly frequent during kitchen activities (88 and 84 occurrences per hour, respectively). While this represents only a subset of \ac{ADLs}, it can be inferred that pick-and-place movements are an essential part of \ac{ADLs}. Many researchers thus focus on different ways to allow people with motor impairments to perform pick-and-place tasks using robots~\citep{Fattal.2019,wang2019towards,Kim.2012,shafti2019gaze}. For an assistive robotic arm, implementing pick-and-place operation as shared control aims to have the user remain in control of this fundamental interaction but use decisions or recommendations made by the robot to improve the usability of the system. There are various ways to accomplish such user-centered shared control.

One of the simplest types of shared control for assistive robot arms is time-optimal mode switching~\citep{Herlant.2016modeswitch}. Time-optimal mode switching uses the same cardinal \ac{DoFs} as manual control but replaces user-initiated mode switches with automatic ones. The results of a user study comparing this type of shared controls -- using a 2D simulated mobile robot and Dijkstra's algorithm to predict when the robot should switch modes -- by \citet{Herlant.2016modeswitch} showed that users prefer time-optimal mode switching to manual controls. A different approach to shared control is the blending of user input and decisions by the autonomous system, for example, to control a powered wheelchair manually while the system helps the user to avoid obstacles \citep{ezeh2017}. In principle, users manually control the robot along cardinal \ac{DoFs}, but the robot interferes to prevent accidents. A similar approach could also be viable for assistive robotic arms. To perform more complex movements using a low number of input \ac{DoFs}, \citet{quere2020} proposed \emph{Shared Control Templates}, a system which \enquote{defines task-specific skills as Finite State Machines in which each phase specifies taskrelevant (sic) input mappings and active constraints}. For example, when pouring water from a bottle, the robot's movement is constrained such that the bottle's opening is always above the container the water is being poured into, but the rotation of the bottle is controlled by user input.
On the more autonomous end of the shared control spectrum, \citet{jain2015} proposed a control method employing a Body-Machine Interface. Using this Body-Machine Interface, users control the execution speed of trajectory segments planned by an autonomous system, switching between these planned segments when appropriate \citep{jain2015}. Thus, only two input-DoFs are required to perform complex tasks while the user remains in control -- one to control movement speed and one to switch segments. 

\begin{figure}
\centering
\includegraphics[width=\linewidth]{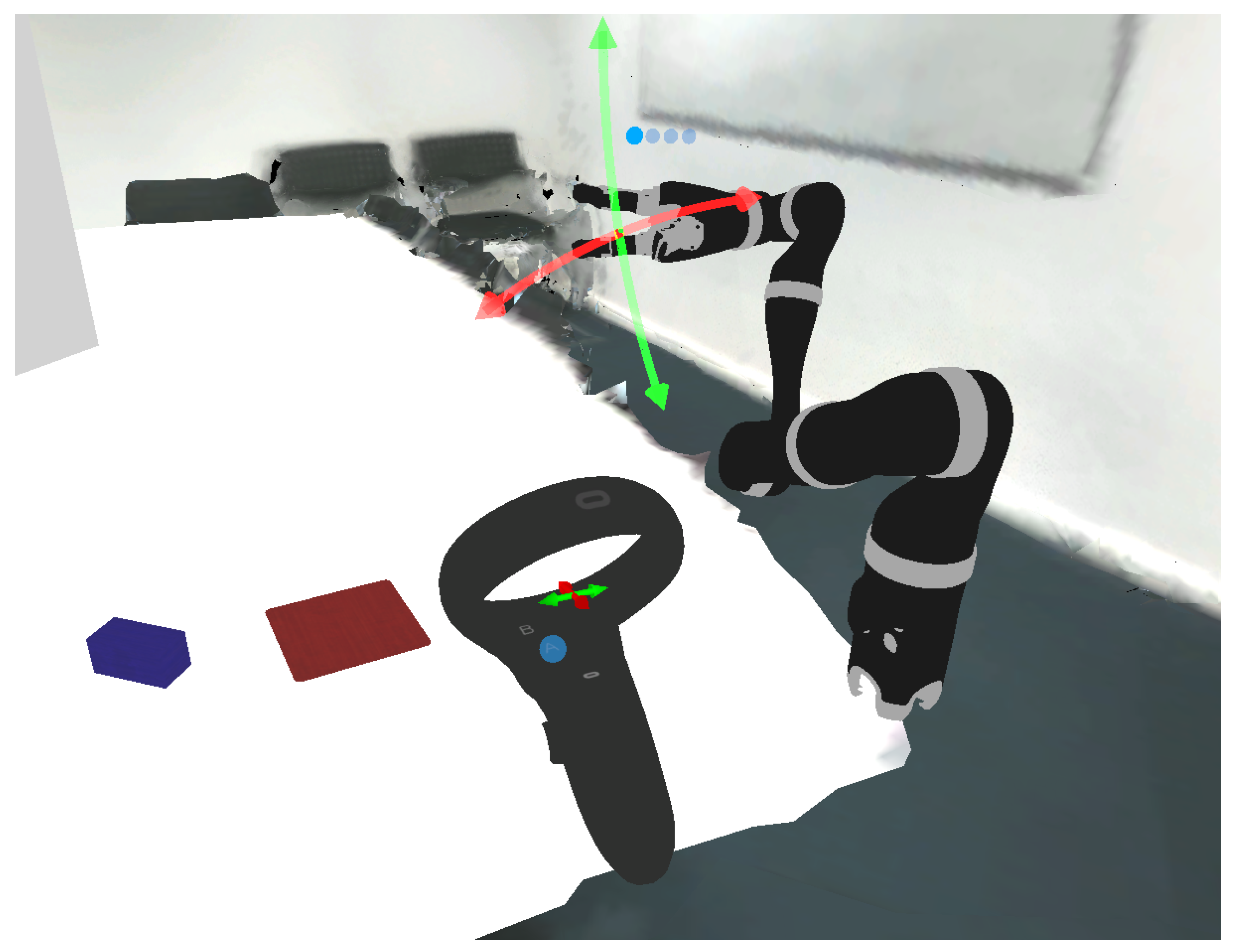}
\centering
\caption{\textbf{Adaptive DoF Mapping Controls \citep{Kronhardt.2022adaptOrPerish}}}
\Description{The figure shows a virtual Kinova Jaco assistive robotic arm augmented with red and green arrows visualizing possible movements. The robot is mounted to a table facing left. Four blue spheres above the robot's gripper indicate the possible modes. In the foreground, an Oculus Quest motion controller augmented with red and green arrows at its control stick indicates the corresponding movements the robots will perform if the control stick is engaged. A blue block and a red surface are visible next to the robot arm.}
\label{fig:admc}
\end{figure}

A fundamentally different approach is the shared control system proposed by \citet{Goldau.2021petra} dubbed \emph{\ac{ADMC}}. The proposed concept works by combining a robotic arm's cardinal \ac{DoFs} according to the current situation and mapping them to a low-DoF input device. This mapping is accomplished by attaching a camera to the robotic arm's gripper and training a \ac{CNN} by having people without motor impairments perform \ac{ADLs} \citep{Goldau.2021petra} -- similar to the learning-by-demonstration approach for autonomous robots \citep{Canal.2016}. In theory, the \ac{CNN} can return a set of DoF mappings ordered by the likelihood of helpfulness given a camera feed of a situation. Compared to previous work, the main advantage of this concept is that users are not restricted to either cardinal \ac{DoFs} or preplanned motions but are given a set of likely movements for a given situation. In addition, the \ac{CNN}-based approach allows the system to be easily extendable in theory, as the same system can be trained to discriminate between many different situations -- making it a viable concept for day-to-day use. \citet{Goldau.2021petra} conducted a small proof-of-concept study with 23 participants comparing the control of a simulated 2D robot with either manual control or \ac{ADMC}. Their results showed that task execution is faster with their \ac{ADMC} but is experienced as more complex by the users \citep{Goldau.2021petra}.

To evaluate the applicability of \citet{Goldau.2021petra}'s work for realistic 3D scenarios, we previously conducted an unsupervised remote experiment in \ac{VR} \citep{Kronhardt.2022adaptOrPerish}. We extended the ideas proposed by \citet{Goldau.2021petra} by applying them to a virtual model of a \emph{Kinova Jaco} (see \autoref{fig:admc}) and heuristically generating the DoF mappings offered to the users -- simulating the output of a \ac{CNN} \citep{Kronhardt.2022adaptOrPerish}. Our results show that while the average number of mode switches significantly decreases when using \ac{ADMC}, neither task completion times nor perceived workload is reduced when directly translating \citet{Goldau.2021petra}'s work to 3D \citep{Kronhardt.2022adaptOrPerish}. Moreover, we identified several problems via thematic analysis of participant reports. The participants reported a perceived lack of predictability of the system regarding the upcoming DoF mappings and the robot's future movement. Another reported issue was the high cognitive demand for understanding the suggested \ac{DoF} mappings. Participants were generally confused by the changing \ac{DoF} mappings compared to manual controls, where \enquote{it was best to intuitively remember where each function was}. We believe these drawbacks arise from the system's novelty and its current complexity level. While \ac{ADMC} remain a promising shared control concept for assistive robot arms, as seen in the reduced number of mode switches, their design needs to be adjusted to address these problems by decreasing the system's overall complexity.

\section{Robot Motion Intent}
\label{sec:background-robotIntents}

Much previous work regarding shared control for assistive robots is focused on the specific interaction between the autonomous system and the human controlling it -- i.e., when is the movement controlled by the autonomous system, and when does the user control it. However, to improve the predictability of the shared control system and thus improve user acceptance, it is necessary to communicate the intended assistance provided by the autonomous system \citep{brooks2020}. Information about a robot's plan and activity can help users better understand and expect a robot's behavior \citep{suzuki2022augmented}. Users generally prefer to have the robot's future movements represented visually \citep{cleaver_2021}. This problem of communicating robot motion intent has been studied extensively with different types of visualization and modalities used~\cite{Pascher.2023robotMotionIntent}. 

\begin{figure}[t]
\centering
\includegraphics[width=0.85\linewidth]{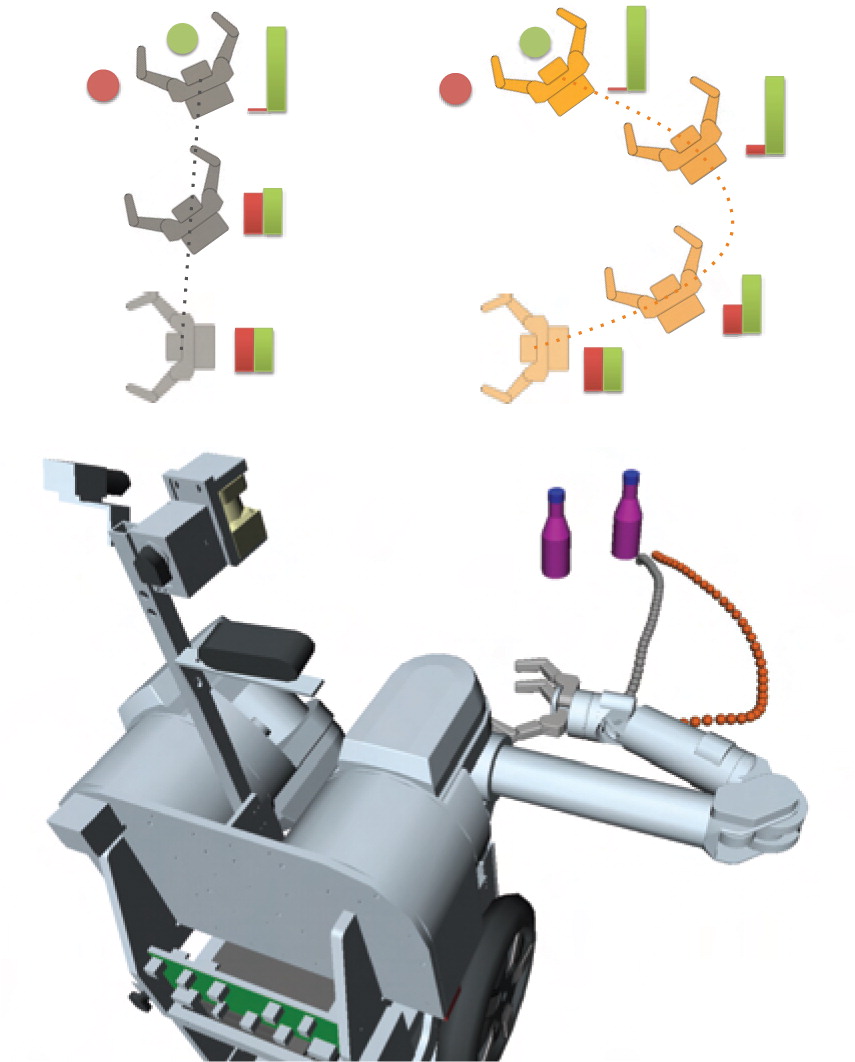}
\centering
\caption[Example of Legible Trajectories]{\textbf{Example of Legible Trajectories}: top left: trajectory optimized for distance, top right: trajectory optimized for legibility, bottom: trajectories in context \citep{dragan2013}}
\Description{Simulated, virtual robotic arm on a mobile platform. The figure shows two different paths of grasping a bottle (optimized for distance, optimized for legibility). Readers can see, that for legible motion the planned trajectory of the robotic arm is longer but easier to understand as oit seems more natural.}
\label{fig:legible-trajectories}
\end{figure}

One solution that does not require additional hardware to communicate robot motion intent is anthropomorphizing the robot. For example, \citet{dragan2013} propose that an assistive robot could communicate its motion intent by selecting a trajectory that may not be the optimal path in regards to distance or execution time (see \autoref{fig:legible-trajectories}) but allows users to understand the robot's intent. Similarly, \citet{gielniak2011generating} designed anthropomorphized robot gestures that elicit anticipation in onlookers. If the robot already has some form of \enquote{eyes}, either as part of its hardware or virtually on a desktop, robot gaze can be used to communicate motion intent \citep{may2015}. 

Robot intent can also be conveyed via different sensory modalities, such as audio or haptics, though current solutions do not explicitly communicate movement direction. \citet{frederiksen2019} used affective audio signals to implicitly convey a soft robot arm's intent to move toward a user. \citet{che2020} used a wearable haptic device to signal robot motion intent to users to avoid collisions.

To convey detailed robot motion intent, researchers often rely on \ac{AR} \citep{walker_2018_armotionintent, ruffaldi2016third, Hetherington_2021, chadalavada2020bi, zein2021deep}, as \enquote{with the help of AR, interaction can become more intuitive and natural to humans}~\citep{makhataeva2020}. \citet{milgram1994taxonomy} define \ac{AR} as \enquote{all cases in which the display of an otherwise real environment is augmented by means of virtual (computer graphic) objects}. There are several different approaches to augmenting real environments, such as \ac{SAR}, Mobile \ac{AR}, and Wearable \ac{AR} \citep{makhataeva2020}. \ac{SAR} augments the environment directly using projection to provide information on surfaces, such as the ground in front of a robot. While \ac{SAR} visualizations are visible to any person in the vicinity, \ac{SAR} use-cases are limited to two-dimensional surfaces in lighting dim enough not to interfere with the projections. Mobile \ac{AR} uses devices such as smartphones or tablets, augmenting the device's camera feed in such a way that objects are registered in 3D space. However, depth perception is difficult due to a lack of stereoscopic information. In contrast, wearable \ac{AR} augments the field-of-view of the user, either by displaying a pass-through of a camera facing the outside of an \ac{HMD} or by using an \ac{HMD} with a transparent screen. As these \ac{HMD}s provide stereoscopic images, virtual objects can be registered at specific 3D positions in the environment and be perceived at the correct depth. 

\subsection{Feedback Modalities for User Attention Guidance}
\label{sec:attention-guidance}
When designing shared control systems, it may be necessary to guide the users' attention to focus on the robot's intended assistance. This guidance is especially necessary if any of the controlling parties -- user or robot -- are moving the robot in a way that may cause collisions or otherwise worsen the situation's outcome. 

To the best of our knowledge, different sensory modalities have not yet been comparatively studied regarding assistive robots and reaction times to their signals. However, there have been extensive comparative studies in a comparable situation: Autonomous driving. When driving autonomously, in systems where the vehicle recognizes a situation it was not programmed or trained to deal with, it will issue a so-called \ac{TOR}. This \ac{TOR} prompts the vehicle's driver to take over control of the vehicle manually to avoid a collision or to drive the vehicle in areas it cannot deal with using its autonomy.

Multiple simulation studies have explored how to effectively communicate \ac{TOR}s and which modalities to use. In general, \ac{TOR}s have been studied using either a single sensory modality (unimodal) or a combination of different sensory modalities (multimodal) simultaneously conveying the \ac{TOR}s. The modalities generally used for \ac{TOR}s are visual, auditory and tactile/haptic. The necessary output devices for these modalities, such as screens, speakers, or vibration motors, are generally included already in many input devices or assistive systems and therefore easily accessible for developers to use. 

Regarding unimodal \ac{TOR}s, \citet{petermeijer2017} conducted a simulation study with 101 participants, comparing visual, auditory, and tactile \ac{TOR} modalities under the influence of visual, auditory, and bimodal (visual-auditory) distractions. They found no effect of the distraction modalities on reaction times regardless of \ac{TOR} modality. However, they found that the \ac{TOR} modality significantly affected reaction times. Tactile and auditory \ac{TOR}s led to significantly faster reaction times than visual \ac{TOR}s and higher self-reported usefulness ratings, indicating that these modalities are better suited to guide user attention when used unimodally.

\citet{PETERMEIJER2017204} also conducted a 24-participant study exploring whether the direction a \ac{TOR} is presented from influenced the participants' approach to avoiding a collision. Comparing auditory, tactile, and auditory-tactile (bimodal) \ac{TOR}s conveyed from the left, right, or both sides, they discovered that the directionality of the \ac{TOR} had no impact on the participants' approach to avoiding a collision. However, they discovered that the bimodal approach showed a faster reaction than the tactile approach and better self-reported usefulness scores \citep{PETERMEIJER2017204}.

This result, indicating that using multiple modalities to communicate \ac{TOR}s improves reaction times, is further supported by the work of \citet{yun2020multimodal}, exploring seven different modality combinations to communicate \ac{TOR}s: visual, auditory, haptic, visual-auditory, auditory-haptic, visual-haptic, and visual-auditory-haptic. The results of their 41-participant simulation study show that unimodal visual \ac{TOR}s lead to significantly higher reaction times than the other approaches and that the visual-auditory-haptic and visual-haptic \ac{TOR}s had the lowest reaction times~\citep{yun2020multimodal}. 

The results of these simulation studies, as well as research regarding reaction times to different sensory stimuli in general~\citep{burke2006comparing, diederich2004bimodal, kosinski2008literature}, indicate that using multimodal feedback should lead to the lowest possible reaction times in shared control systems. This information is essential for guiding users attention towards intended assistance and may improve the effectiveness of shared control methods overall. 

\section{Discussion \& Conclusion}
In this paper, we provided an overview for the motivation of developing shared control systems for assistive robotic arms. Contrary to the dominant goal of \ac{HRI}, which is to reach increased autonomy of the robot and rely on shared control for those situations where human skills remain superior, for assistive robotic arms, a main objective is to increase the self-determination and physical autonomy in peoples' life. Research has shown that a high degree of robot autonomy can be counter-productive. For effective shared control systems, however, it is important that robot support during operation is both legible and predictable to the user. Research has shown that addressing the basic task of pick-and-place operation can be a promising starting point for shared control due to its high occurrence, with the ultimate goal being to relieve the manual control of the \ac{DoF} complexity of a robotic arm. 

Through related work and our own research, we identified three main aspects that need to be considered for shared control systems in the context of robotic arms:
\begin{itemize}
    \item \textit{Less is More}: While the goal is to keep users in control, the complexity of both the robot interaction and the \ac{DoF} limitations for available input devices can easily make the system very difficult to use. Therefore, in order to find the sweet spot for shared control, we propose to start with a rather minimized set of user interaction and increase that on demand and depending on the individual capabilities. Users should not be confused by too many interaction options or overly complex movements. While optimal ways of accomplishing a goal may require complex intervention from the robot, these interventions may be difficult for users to understand, and therefore trust. In addition, keep the \ac{DoF} of input devices low as this maximizes the amount of assistive devices capable of controlling the robot.

    \item \textit{Pick-and-Place Matters}: Our previous research and related work show that pick-and-place tasks are ubiquitous and necessary to perform \ac{ADLs}. It is, therefore, important that shared control is first implemented for these simple tasks before more complex sequences are examined. If users struggle to understand shared controls for pick-and-place tasks, we believe it is highly likely that more complex tasks may cause further frustration.
    
    \item \textit{Communicating Intent}: While the robot system may have the interests of the user at heart, the user still needs to build trust, which requires a level of transparency and legibility the user can comprehend. They should also be able to interfere with the robot's part of the control, in case the robot makes a mistake or gives inappropriate suggestions for interaction. Communicating intent further requires having the user pay attention or guiding the attention of the user, requiring multi-modal stimuli, depending on the situation and the capabilities of the user.
\end{itemize}

\bibliographystyle{ACM-Reference-Format}
\bibliography{sample-base}


\end{document}